\def\cm2{cm$^{-2}$}

\def\c2{C~{\sc ii}}
\def\c4{C~{\sc iv}}
\def\fe2{Fe~{\sc ii}}
\def\fe3{Fe~{\sc iii}}
\def\mg1{Mg~{\sc i}}
\def\mg2{Mg~{\sc ii}}
\def\si2{Si~{\sc ii}}
\def\si4{Si~{\sc iv}}
\def\al2{Al~{\sc ii}}
\def\al3{Al~{\sc iii}}
\def\o1{O~{\sc i}}
\def\n1{N~{\sc i}}
\def\h1{H~{\sc i}}

\def\approxlt{\mathrel{\spose{\lower 3pt\hbox{$\sim$}}
        \raise 2.0pt\hbox{$<$}}}
\def\approxgt{\mathrel{\spose{\lower 3pt\hbox{$\sim$}}
        \raise 2.0pt\hbox{$>$}}}

\documentclass[article]{has40}
\usepackage{graphicx}
\usepackage{amssymb}
\tabletypesize{\normalsize}  % AMcW

\def\plotone#1{\centering \leavevmode
\includegraphics[width=.95\columnwidth]{#1}}

\def\plottwo#1#2{\centering \leavevmode
\includegraphics[width=.45\columnwidth]{#1} \hfil
\includegraphics[width=.45\columnwidth]{#2}}

\def\plotone#1{\centering \leavevmode
\includegraphics[width=.95\columnwidth]{#1}}

\def\plottwo#1#2{\centering \leavevmode
\includegraphics[width=.45\columnwidth]{#1} \hfil
\includegraphics[width=.45\columnwidth]{#2}}

\shortauthors{Wehrung \& Layden}
\shorttitle{Bright Variables in M5}

\begin{document}
\large    %AMcW  The conference proceedings will employ large size print
\pagenumbering{arabic}
\setcounter{page}{202}

\title{Long Period Variable Stars in the Globular Cluster M5}

%
% Here is an example of how to include the author names and affilitations
%
\author{{\noindent Michael Wehrung and Andrew Layden\\
\\
{\it Physics \& Astronomy Dept., Bowling Green State Univ., Bowling Green, OH, USA} 
%\author{{\noindent Michael Wehrung{$^{\rm 1}$} and Andrew Layden{$^{\rm 1}$}\\
%\\
%{\it (1) Physics \& Astronomy Dept., Bowling Green State Univ., Bowling Green, OH, USA} 
}
}

%
% And here is how to add the e-mail addresses
%
\email{laydena@bgsu.edu}

% If you reall you need to add an alternate institution, then update and uncomment the following line.
% It's not very pretty though
%\altaffiltext{}{(3) also at The Insitute for the Insane}

\begin{abstract}
We report results of $VI$ time series photometry for the bright
globular cluster M5 taken between 2007-2011 using the 0.5-m telescope
at Bowling Green State Univ. and the PROMPT \#4 telescope at Cerro
Tololo.  We used DAOPHOT to obtain photometry to a limiting magnitude
of $V \approx 19$ mag.  A search for variable stars using the
DAOMASTER variability index enabled us to recover the three known
bright variables (two Cepheids and one red, long period variable, or
LPV) and many known RR Lyrae stars, and to discover as many as
thirteen new, low-amplitude LPVs.  We present light curves of several
stars and analyze periods and amplitudes where applicable.
\end{abstract}

\section{Introduction}
The globular cluster M5 (NGC~5904) is a bright, well-studied cluster
of moderate metallicity. In addition to its abundant RR Lyrae stars,
M5 has two well-studied type II Cepheids, V42 and V84, which are the
subject of recent CCD photometry by Rabidoux et al. (2010). The irregular LPV star V50 has been known since Bailey
(1917) but has never been the subject of CCD photometry. No other
bright variables are known in M5.

The goals of our CCD study are to seek new long period variables and
obtain their light curves (LCs), amplitudes, and periods; to compare
them with metal-rich LPVs in clusters and the LMC; and to test the
hypothesis that period and amplitude increase as stars evolve up the
giant branch.

\section{Data and Analysis}
Images were obtained in $VI$ using the BGSU 0.5-m and Apogee Ap6e CCD
during the summers of 2007, 2009, and 2010, and using the PROMPT \#4
0.4-m and Apogee Alta CCD at CTIO in the fall of 2010 and throughout
2011 via the robotic SKYNET client.  Standard image processing was
performed on each image. We ran the images through the DAOPHOT and
ALLFRAME point-spread function fitting photometry packages (Stetson
1987, 1994), and used Stetson's DAOMASTER program to combine data
across images and search for variable stars.  Photometric calibration
was performed using in-field standards from Stetson (2000). To obtain
a more complete census of bright variables in the cluster center, we
employed the ISIS image subtraction program (Alard 2000).

Figure 1 shows color-magnitude diagrams of stars in the BGSU data
located more than 1 arcmin from the center of M5, along with magnitude
as a function of variability index, $\Lambda$.  The two stars with
large $\Lambda$ are the known Cepheids V42 and V84. RR Lyrae stars at
the level of the horizontal branch also show large $\Lambda$ but these
stars have been studied well by other observers, so we do not consider
them further.  Some stars at the tip of the giant branch have large
$\Lambda$ and are candidate LPV stars.

%Figure 1: CMD
\begin{figure*}
\centering
\plotone{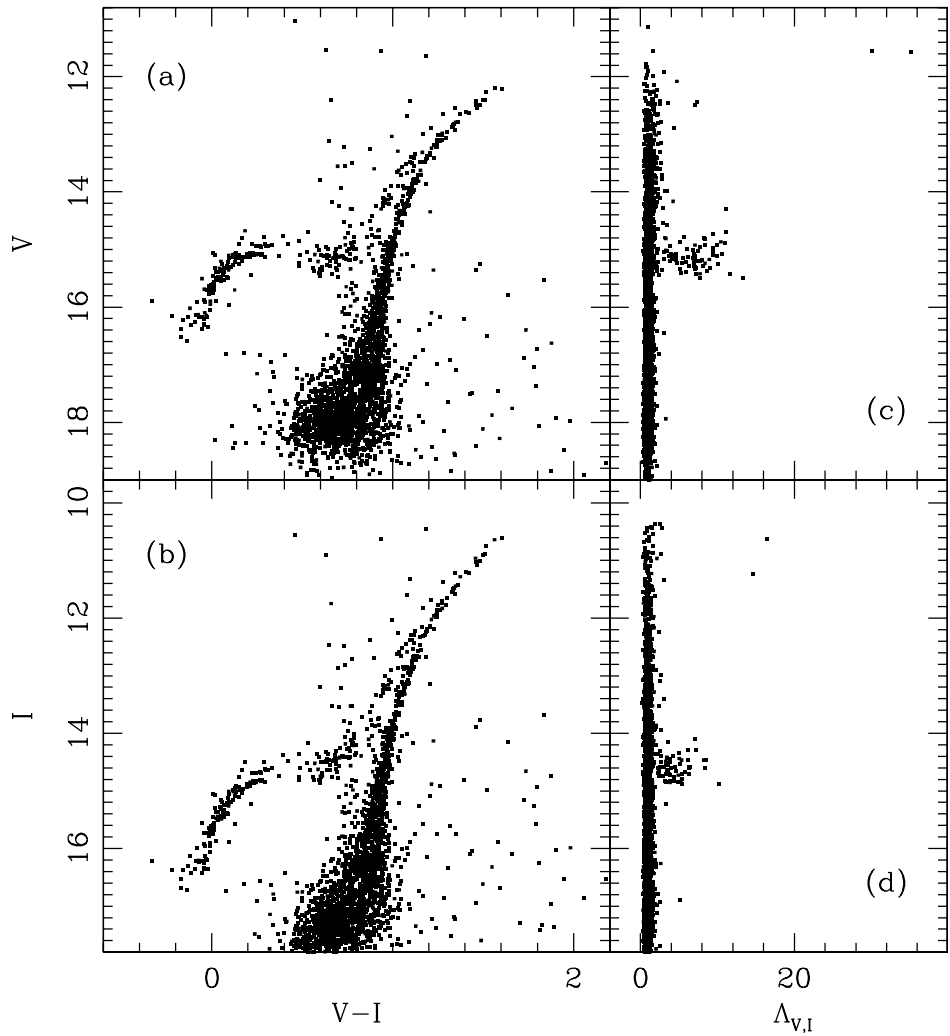}
\vskip0pt
\caption{The color-magnitude diagrams and variability indices of stars in M5. }
\label{cmd}
\end{figure*}

\section{Variable Stars}
Figure 2 shows the light curves of type-II Cepheids V42 and V84. We
folded V42 using the period from Rabidoux et al. (2010), suggesting
little period change since their 2003-6 observations. The scatter at
$\phi \approx 0.25$ is real. It was more difficult to characterize the
light curve of V84. We folded the PROMPT data using the Rabidoux
period of 26.93 d, but a shorter period near 26.74 d was needed to
fold the earlier BGSU data. We are collaborating with Horace Smith to
use these and new MSU data to extend the period change analysis from
Rabidoux et al. (2010).

%Figure 2: Light curves in $V$ (blue) and $I$ (red).  In all LCs, data from images with better/worse than median seeing are shown with open circles and crosses, respectively. 
\begin{figure*}
\centering
%\includegraphics[width=12cm]{mgsio2.eps}
%\plotone{vartest500.eps}
\plottwo{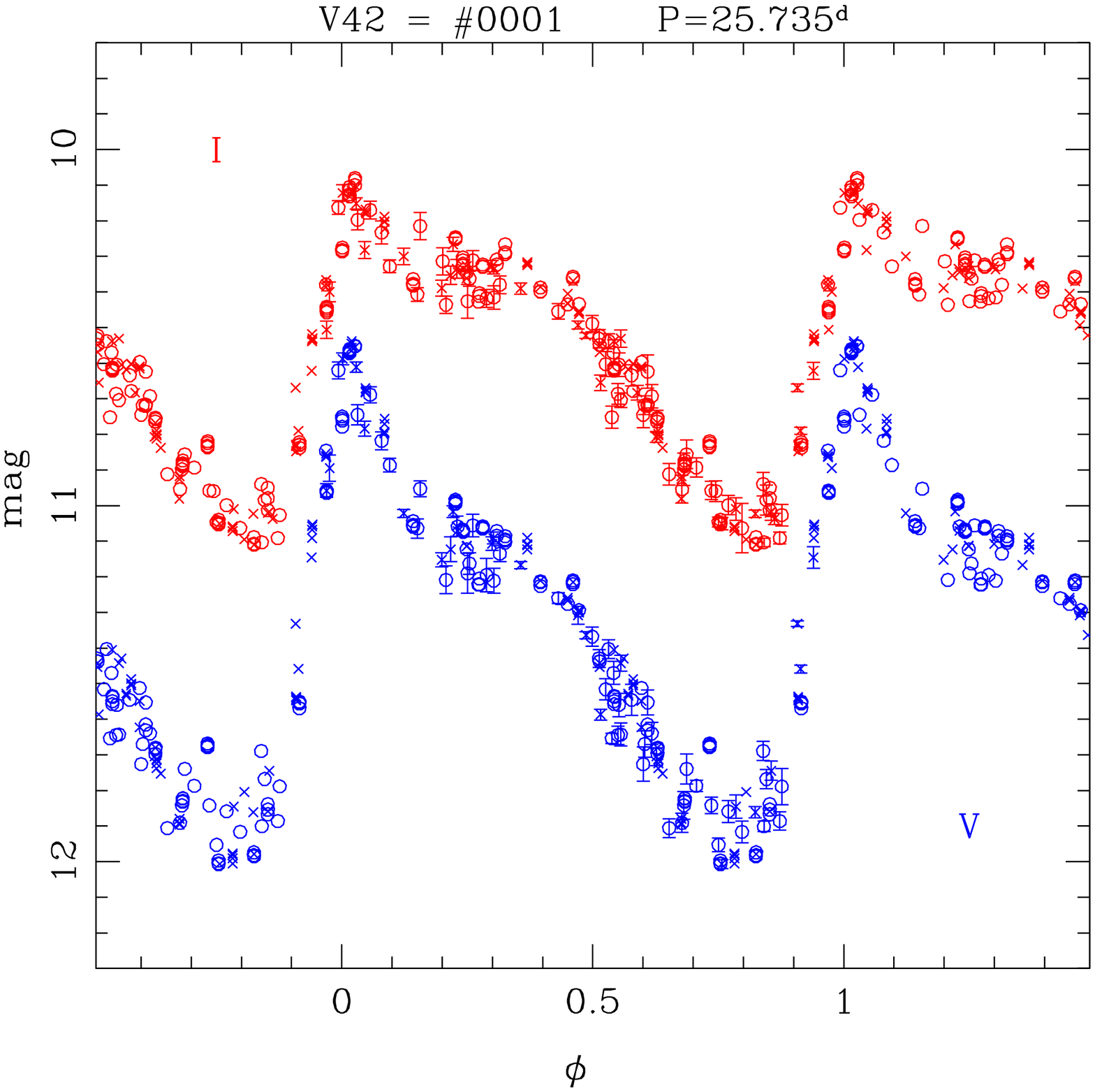}{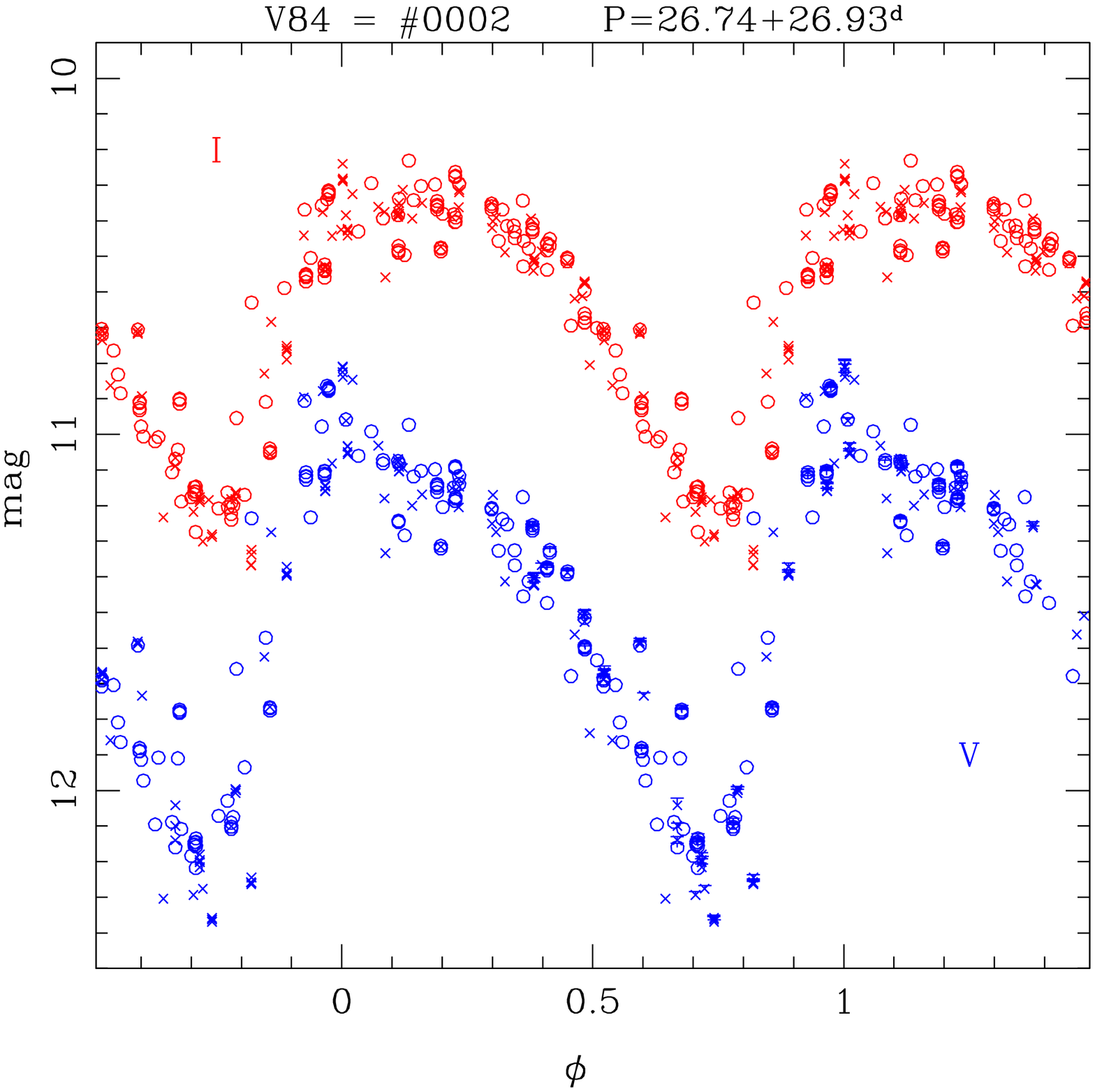}
\vskip0pt
\caption{ Light curves in $V$ (blue) and $I$ (red).  In all LCs, data
from images with better/worse than median seeing are shown with open
circles and crosses, respectively. }
\label{cepheid_lcs}
\end{figure*}

In Figure 3, we show the light curves of two irregular LPV stars, the
known variable V50 and the newly detected variable ID\#4.  Notice how
the amplitude of both stars waxes and wanes, as if two oscillations
with similar periods were beating against each other. From his
photographic data, Bailey (1917) estimated the period of V50 to be 106
days. We reanalyzed his data using the phase dispersion minimization
(PDM) method, and found a period of $105 \pm 2$ days. Using PDM on our
CCD data, we obtained a period of $102 \pm 1$ days. However, the
photographic data of Coutts Clement \& Sawyer Hogg (1977), taken from
1946-1957, shows no coherent period in the light curve or via
PDM. More observations over decades may be required to establish
whether the underlying periodicity is fundamentally stable.

We estimated periods and photometric ranges ($V_{min} - V_{max}$) for
each of the thirteen LPV stars. Several of the stars seem to show very
short period variations ($<30$ day cycles) on a long, secondary
period, seen in $\sim$30\% of irregular LPVs (Kiss et al. 1999).

%Figure 3: Light curves for the two LPVs with the highest variability indices.
\begin{figure*}
\centering
%\includegraphics[width=12cm]{mgsio2.eps}
%\plotone{vartest500.eps}
\plottwo{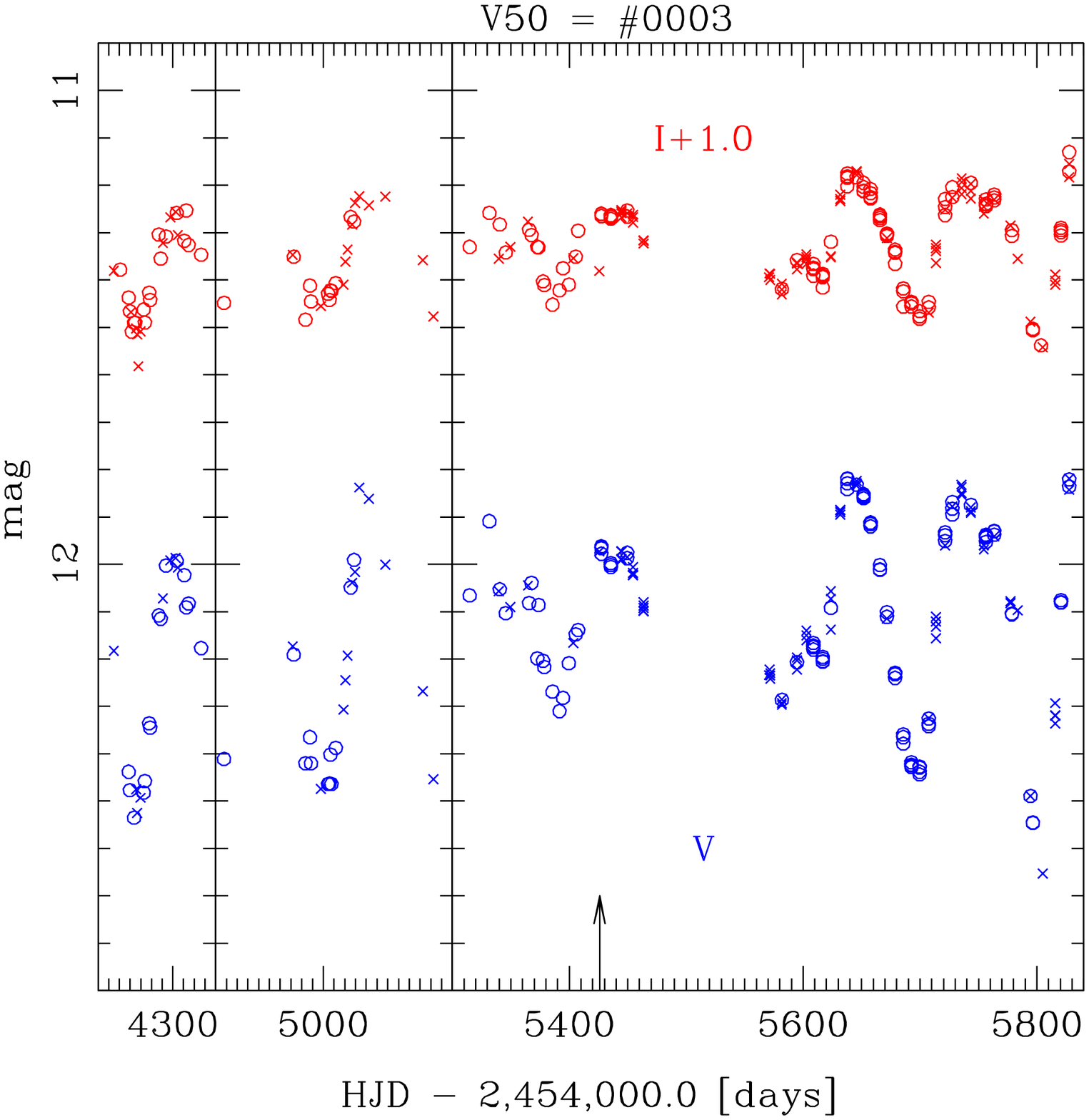}{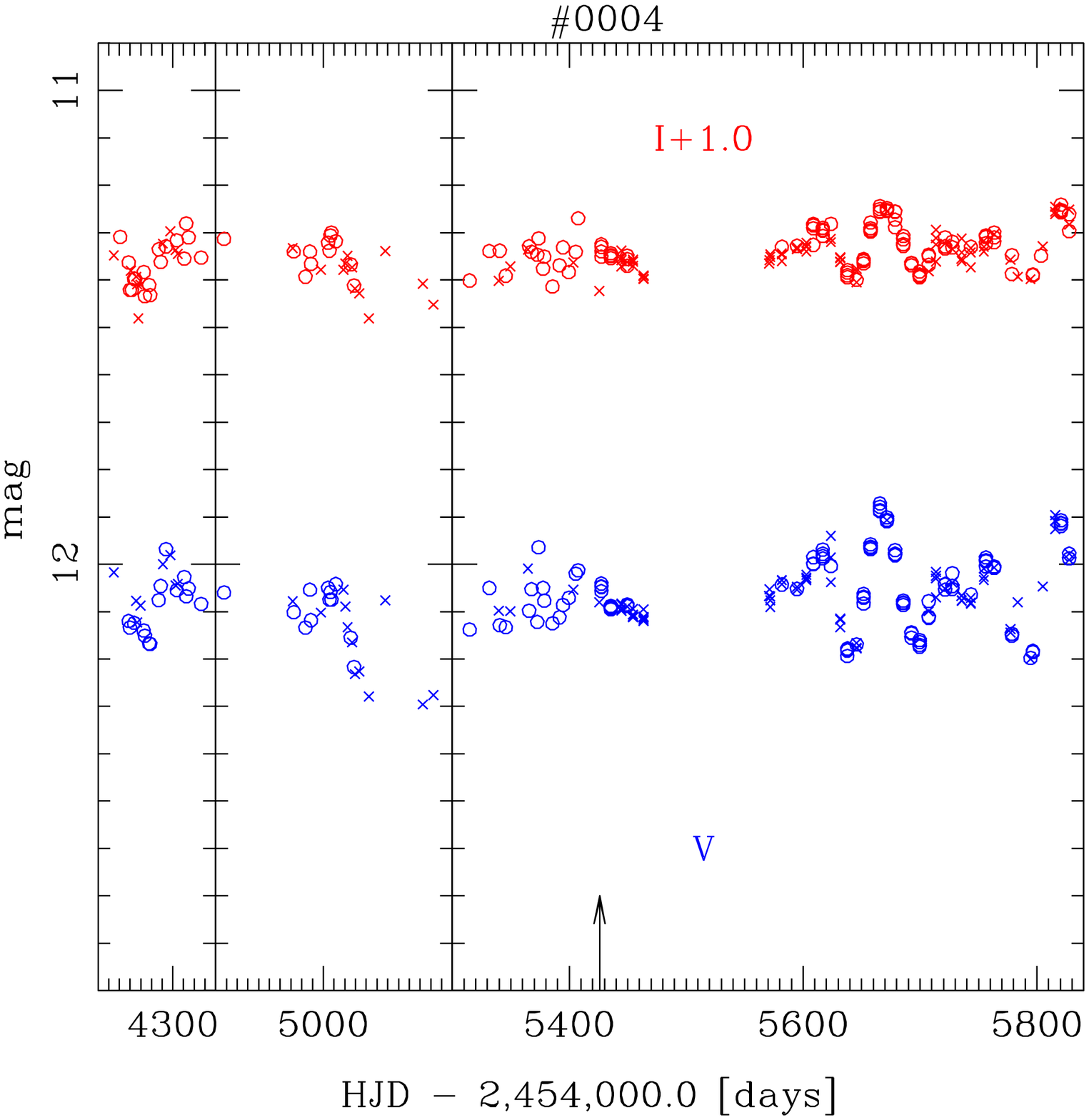}
\vskip0pt
\caption{ Light curves for the two LPVs with the highest variability indices. }
\label{cepheid_lcs}
\end{figure*}

Figure 4 shows the LPVs' $V-I$ color, magnitude range, and period
plotted against $V$ magnitude. The red giant stars in M5 show no sign
of variability at $V \approx 1$ mag below the giant branch tip, but
some stars show $ = 0.1$ to 0.2 mag variations on $< 30$ d timescales
as much as $V \approx 0.5$ mag below the tip. By $V \approx 0.2$ mag
below the tip, variations become more prominent, culminating with
V50. Star \#18 is a notable exception to this pattern.  

%\vspace*{-2.0cm}

\begin{figure*}
\centering
\plotone{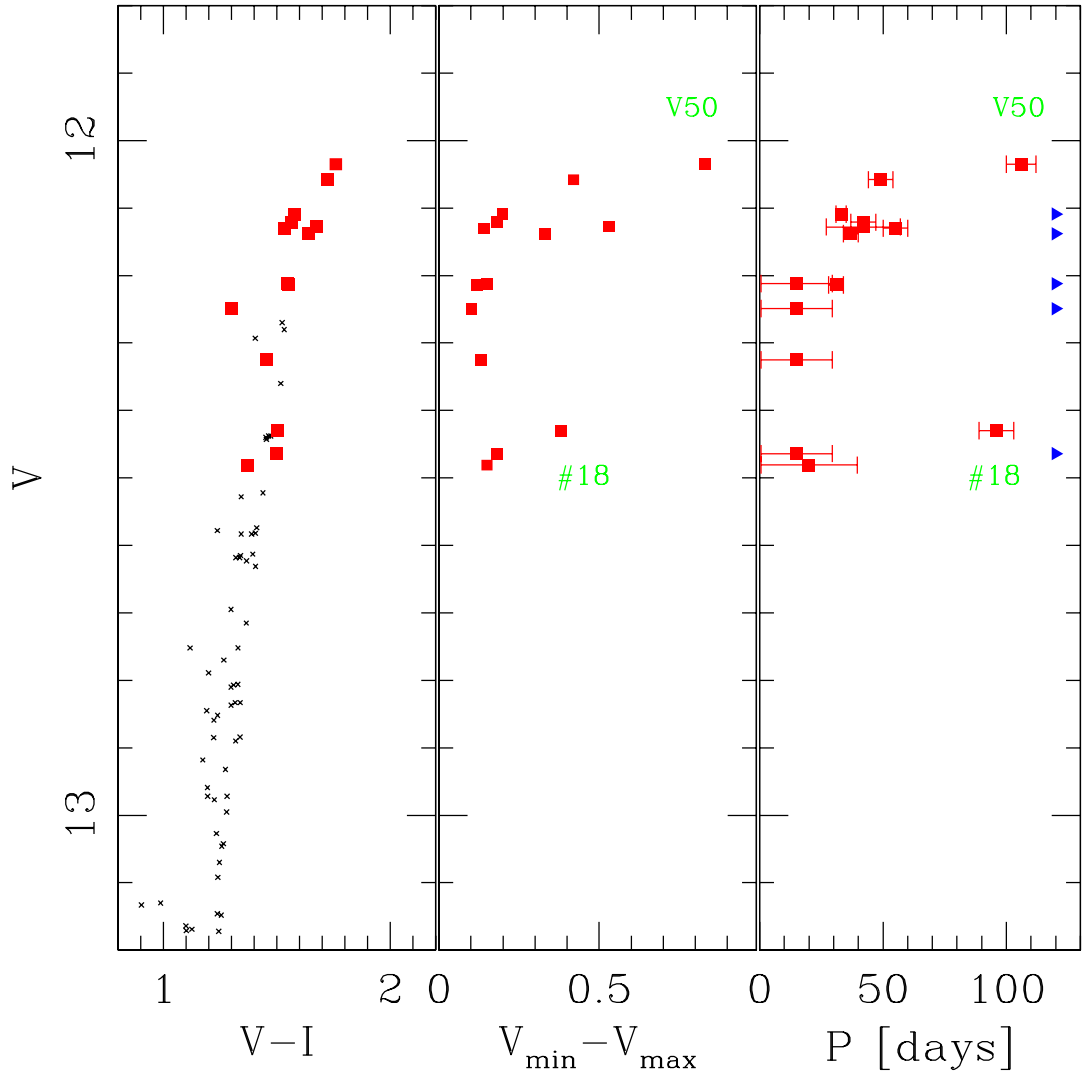}
%\plottwo{wehrung_f2a.eps,wehrung_f2b.eps}
\vskip0pt
\caption{ ($Left$) The CMD zoomed in on the red giant stars; variable
stars are marked with red squares. ($Center$) Periods, and ($right$)
magnitude ranges of the variables are shown. Error bars show the
variation in amplitude from cycle to cycle, while blue triangles mark
stars with possible long secondary periods. }
\label{per_ampl}
\end{figure*}

%\vspace{1.0cm]

~

~

\section{Conclusions}
We present the first time series CCD photometry of the LPVs in M5. We
recovered the known irregular variable V50 and discovered thirteen
new, low-amplitude, irregular LPVs. The pulsation characteristics,
period and magnitude range, increase with position up the giant
branch, suggesting that stars become increasingly unstable to
pulsation as they evolve toward cooler surface temperatures and lower
surface gravities. We will investigate the exception to this rule,
ID\#18, checking whether it is a cluster member using existing proper
motion and radial velocity surveys.  Compared to LPVs in more
metal-rich systems, the M5 LPVs tend to have lower amplitudes, shorter
periods, and less regular cycles, no doubt because the bluer giant
branch of M5 reaches less far into the Mira instability domain.

~

Acknowledgements: The authors acknowledge support from BGSU's
SETGO Summer Research program, which is funded by the NSF.

\end{document}